\documentclass[preprint2]{aastex}
\usepackage{hyperref,amsmath,natbib}
\newcommand{\kms}{{\ensuremath{\mathrm{~km~s^{-1}~}}}}

\begin{document}

\begin{abstract}
Spiral galaxies must acquire gas to maintain their observed level of star formation beyond the next few billion years \citep{2008AJ....136.2782L}. A source of this material may be the gas that resides between galaxies, but our understanding of the state and distribution of this gas is incomplete \citep{2012ApJ...759...23S}. Radio observations \citep{2004A&A...417..421B} of the Local Group of galaxies have revealed hydrogen gas extending from the disk of the galaxy M31 at least halfway to M33. This feature has been interpreted to be the neutral component of a condensing intergalactic filament \citep{2001ApJ...552..473D} which would be able to fuel star formation in M31 and M33, but simulations suggest that such a feature could also result from an interaction between both galaxies within the past few billion years \citep{2008MNRAS.390L..24B}. Here we report radio observations showing that about 50 per cent percent of this gas is composed of clouds, while the rest is distributed in an extended, diffuse component. The clouds have velocities comparable to those of M31 and M33, and have properties suggesting they are unrelated to other Local Group objects. We conclude that the clouds are likely to be transient condensations of gas embedded in an intergalactic filament and are therefore a potential source of fuel for future star formation in M31 and M33.
\end{abstract}
\title{Discrete clouds of neutral gas between the galaxies M31 and M33}
	\author{Spencer A. Wolfe \altaffilmark{1}, D.J. Pisano \altaffilmark{1,2}, Felix J. Lockman \altaffilmark{3,4}, Stacy S. McGaugh \altaffilmark{5}}
	\vspace{-10.0em}
	\author{Edward J. Shaya \altaffilmark{6}}
   \altaffiltext{1}{Department of Physics, West Virginia University, P.O. Box 6315, Morgantown, West Virginia 26506}
   \altaffiltext{2}{Adjunct Assistant Astronomer at the National Radio Astronomy Observatory}      
   \altaffiltext{3}{National Radio Astronomy Observatory, P.O. Box 2, Green Bank, West Virginia, 24944}
   \altaffiltext{4}{Adjunct Professor, Department of Physics, West Virginia University}
    \altaffiltext{5}{Department of Astronomy, Case Western Reserve University, Cleveland, Ohio, 44106}
   \altaffiltext{6}{Department of Astronomy, University of Maryland, College Park, Maryland, 20742}
\maketitle

A recent study \citep{2012AJ....144...52L} of the region between M31 and M33 using the Robert C. Byrd Green Bank Telescope (GBT) at $9'$ angular resolution has confirmed the presence of neutral hydrogen at levels of a few times $10^{17}$ cm$^{-2}$. The study also showed that this neutral gas has a velocity similar to the systemic velocities of M31 and M33, and is thus not the product of confusion with foreground emission from the Milky Way. The sensitivities of the observations were, however, insufficient to provide a detailed understanding of the structure and extent of this gas. Observations were needed with high angular resolution, high velocity resolution and high sensitivity to distinguish clouds from diffuse gas, and an intergalactic filament from tidal debris. Determining the origin of the gas is important, for if it is from an intergalactic filament, it would represent a new source of material to fuel future star formation in M31 and M33.

\begin{figure*}
\epsscale{1.4}
\plotone{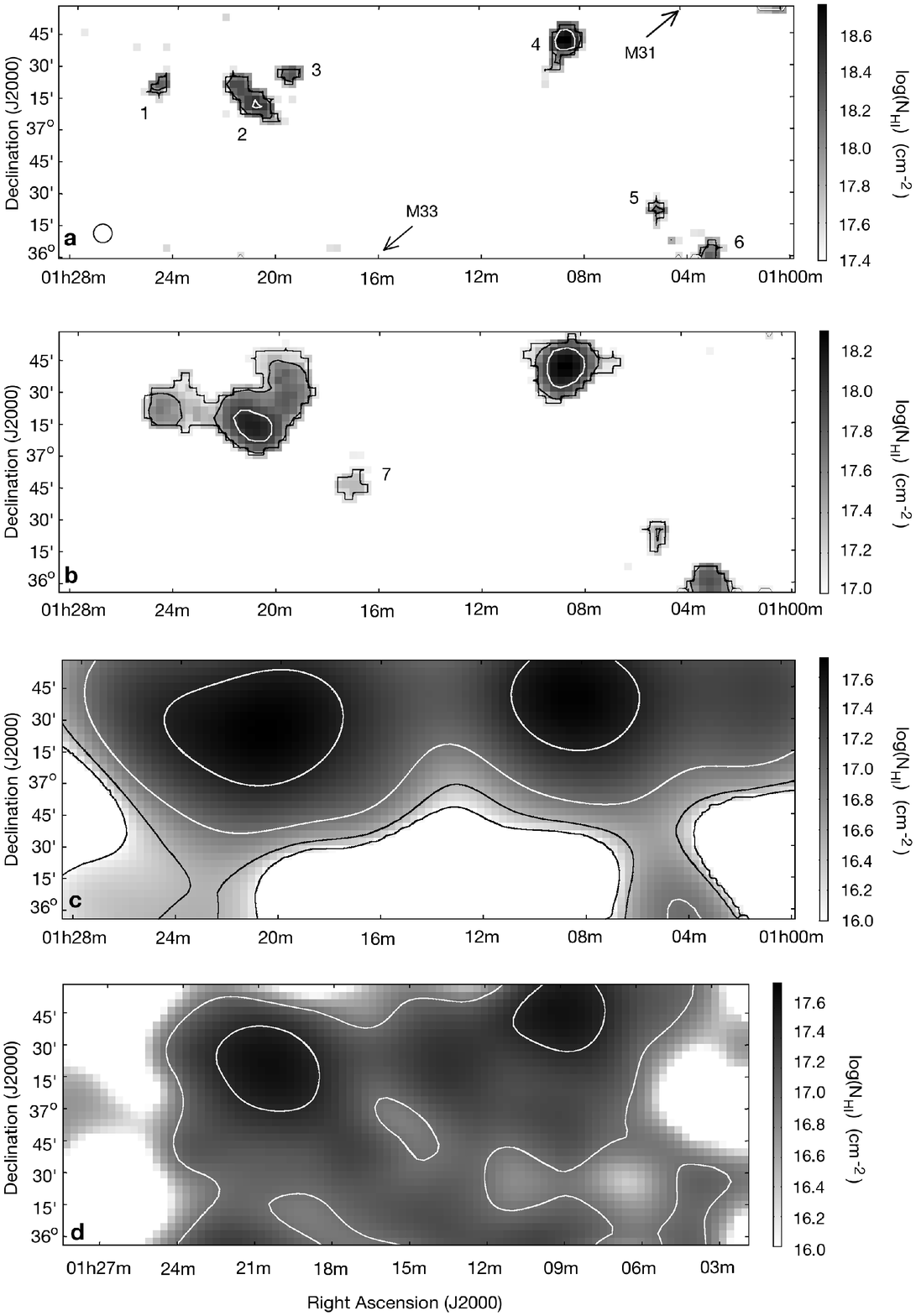}
\caption{{\bf HI column density maps.} Maps of 21 cm emission over a 12 square-degree region between M31 and M33 \citep{2008AJ....135.1983C, 2011A&A...536A..81B}. The data from which these plots were compiled are available in the Supplementary Information. {\bf a}, Column density map of HI (in units of cm$^{-2}$) of the section of the area between M31 and M33 mapped with the GBT. This is a result of nearly 250 hours of integration. The 3$\sigma$ sensitivity limit in N$_{\rm HI}$ is $\rm{2.7\times10^{17}~cm^{-2}}$ for a linewidth of 25\kms at $9'$ angular resolution with a velocity resolution of 5\kms. The contours are for $\log(\rm N_{HI})=17.5, 18.0, 18.5$. There are six confirmed HI clouds, labeled in order of decreasing right ascension. M31 is to the northwest and M33 to the southeast, as indicated by the arrows. Both galaxies are about five degrees from the respective edges of the map. {\bf b}, The GBT data, smoothed to a $15'$ resolution to match the sensitivity of the WSRT data ($\rm{1\times10^{17}~cm^{-2}}$). A seventh cloud is now visible, labeled 7. The contours are for $\log(\rm N_{HI})=17.0, 17.5, 18.0$. {\bf c}, The GBT data, smoothed to a $49'$ resolution to match the angular resolution of the WSRT data. The contours are for $\log(\rm N_{HI})=16.0, 16.5, 17.0, 17.5$. {\bf d}, Part of the WSRT HI survey \citep{2004A&A...417..421B} that first detected neutral gas between M31 and M33, at an angular resolution of $49'$ and a velocity resolution of 18\kms. HI emission was detected over a wide area at ${\rm N}_{\rm HI}\gtrsim10^{17}$ cm$^{-2}$, most of it only 2-3 times the noise level. The contours are for $\log(\rm N_{HI})=17.0, 17.5$.}
\end{figure*}

\renewcommand{\tabcolsep}{4pt}
\begin{table*}
\centering
\centerline{
\begin{tabular}{|c|c|c|c|c|c|c|c|c|}
\hline
Cloud & R.A. & DEC & LSR Velocity & FWHM & T$\rm{_{B, Peak}}$ & $\log$(N$\rm{_{HI})_{Peak}}$ & M$\rm{_{HI}}$ & Diameter \\
... & H:M:S & D:M:S & \kms & \kms & mK & $\rm{cm^{-2}}$ & M$_{\odot}$ & ($\pm 1$) kpc \\
\hline
1 & 01:24:42 & +37:23:02 & -297$\pm$1 & 23$\pm$3 & 47$\pm$4 & 17.6$\pm$0.04 & (1.2$\pm$0.2)$\times$10$^{5}$ & $\leq$ 2.8 \\
2 & 01:20:52 & +37:16:58 & -236$\pm$1 & 28$\pm$2 & 69$\pm$4 & 17.8$\pm$0.03 & (3.5$\pm$0.3)$\times$10$^{5}$ & 6.4 \\
3 & 01:19:25 & +37:31:12 & -226$\pm$2 & 28$\pm$4 & 41$\pm$4 & 17.6$\pm$0.04 & (1.1$\pm$0.2)$\times$10$^{5}$ & $\leq$ 2.4 \\
4 & 01:08:30 & +37:44:51 & -277$\pm$1 & 34$\pm$3 & 91$\pm$4 & 17.9$\pm$0.02 & (4.0$\pm$0.3)$\times$10$^{5}$ & 4.8 \\
5 & 01:05:09 & +36:23:28 & -209$\pm$4 & 26$\pm$3 & 34$\pm$4 & 17.5$\pm$0.05 & (4.2$\pm$1.0)$\times$10$^{4}$ & $\leq$ 2.4 \\
6 & 01:03:10 & +36:01:46 & -282$\pm$4 & 29$\pm$3 & 39$\pm$4 & 17.6$\pm$0.05 & (1.4$\pm$0.2)$\times$10$^{5}$ & $\leq$ 3.2 \\
7 & 01:17:02 & +36:49:34 & -309$\pm$2 & 24$\pm$4 & 10$\pm$2 & 17.0$\pm$0.1 & (4.0$\pm$1.0)$\times$10$^{4}$ & $\leq$ 3.4 \\
\hline
\end{tabular}
}
\caption{{\bf A listing of the HI cloud properties}. R.A. and DEC are respectively the right ascension and declination of each cloud; LSR, Local Standard of Rest; FWHM, full-width at half-maximum of the HI line; T$\rm{_{B, Peak}}$, peak brightness temperature; $\rm{N_{HI, Peak} = 1.82 \times 10^{18} ~ T_{B, Peak} ~ \Delta v}$ is the peak column density, where $\Delta$v $\approx 5$ \kms; M$\rm{_{HI}}$ is the HI mass of the cloud, in solar masses, measured to the 3$\sigma$ column density contour in the maps; diameter is measured to the same contour along the major axis of the cloud, if one is apparent. The $\pm$1 kpc error in the diameter is due to the pixel size of our maps projected to a distance of $\sim$ 800 kpc. The values for cloud 7 were obtained from the $15'$ map. Quantities were calculated assuming a distance to the clouds of 800 kpc \citep{2012AJ....144...52L}.}
\end{table*}

\begin{figure}
\epsscale{1.0}
\plotone{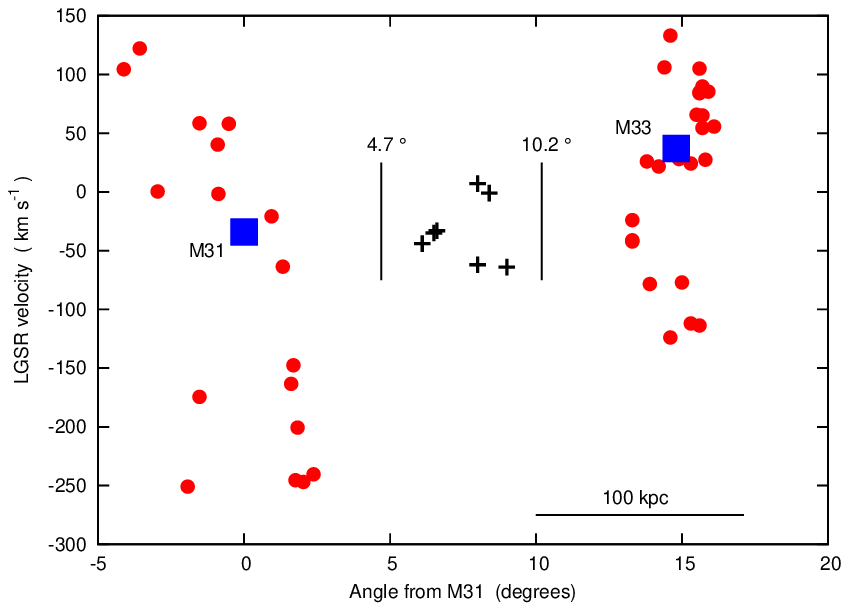}
\caption{{\bf A position-velocity plot of the HI clouds and selected Local Group objects.} The local group standard of rest (LGSR) velocities were calculated using the NASA Extragalactic Database. Large blue squares indicate M31 and M33, both labeled. Filled red circles are the high velocity clouds of each galaxy \cite{2004ApJ...601L..39T, 2008A&A...487..161G, 2008MNRAS.390.1691W, 2009ApJ...703.1486P}. Crosses show clouds detected in the GBT maps; the vertical bars mark the GBT map boundaries, in terms of angle from M31. The HVCs lie within $\sim50$ kpc of each galaxy, whereas new clouds are found $\gtrsim$ 100 kpc from both galaxies. The clouds also have an inter-cloud velocity dispersion of 25\kms, compared to the inter-cloud velocity dispersion of 70-130 \kms for the HVCs.}
\end{figure}

We have mapped a portion of the hydrogen gas feature between M31 and M33 with improved sensitivity and higher angular and velocity resolution using the GBT. We detect seven clouds (Figure 1); Table 1 lists their properties. The positions and velocities of the brightest clouds are consistent with previous GBT results \citep{2012AJ....144...52L}. The primary difference between the results from the Westerbork Synthesis Radio Telescope (WSRT) survey \citep{2004A&A...417..421B}, which used that array as a collection of single dishes, and our high resolution data is that here the HI emission is resolved into features approximately the size of a GBT beamwidth or smaller, and at much higher column densities. The GBT data convolved to $49'$ angular resolution reproduce the same basic morphology as the WSRT data. The total HI mass detected is $\rm{(1.2\pm 0.1) \times10^{6}~M_{\odot}}$ above a level of $\rm{3.0\times10^{17}~cm^{-2}}$, $\rm{(1.3\pm 0.1) \times10^{6}~M_{\odot}}$ above $\rm{1.0\times10^{17}~cm^{-2}}$ and $\rm{(1.9 \pm 0.1)}$ $\rm{\times10^{6}~M_{\odot}}$ above $\rm{1.0\times10^{16}~cm^{-2}}$ (here $\rm{M_{\odot}}$ is the solar mass). These values do not appear to be consistent with the relationship between the mass of neutral atomic hydrogen ($\rm{M_{HI}}$) and its column density ($\rm{N_{HI}}$) predicted by simulations of intergalactic HI at low redshift \citep{2009A&A...504...15P}. The total HI mass measured across the entire map is $\rm{(2.6\pm 0.2) \times10^{6}~M_{\odot}}$, identical to the total HI mass from the WSRT data over the same area. There is probably additional mass present in the form of ionized gas, as at such low column densities photoionization of the HI from the extragalactic radiation field becomes significant \citep{1993ApJ...414...41M}. 

About 50\% of the neutral hydrogen is in clouds, whose diameters range from $\leq$ 2.4 kpc to 6.4 kpc. This is the size range of dwarf galaxies; however, although stellar surveys \citep{2007ApJ...671.1591I, 2009Natur.461...66M} have detected dwarf galaxies around M31 as well as potential dwarf galaxy remnants, there are no reports of stellar over-densities at the positions of these clouds \citep{2009Natur.461...66M}, though there is no lack of stellar streams in the general vicinity \citep{2013ApJ...763....4L}. The clouds lie much farther from M31 and M33 than each galaxy's respective high velocity cloud (HVC) population and also lie closer to the systemic velocities of M31 and M33 than the HVCs (Figure 2). The clouds are also kinematically cooler as a population than the HVCs, with a lower inter-cloud velocity dispersion by a factor of two to five.

We estimate the baryonic mass and circular rotation speed about the cloud center for the clouds as well as for the HVCs and dwarf galaxies of M31 and M33, and plot them with the baryonic Tully-Fisher relation \citep{1977A&A....54..661T, 2000ApJ...533L..99M} (Figure 3). This allows us to compare the clouds to other members of the Local Group. The clouds are offset from the Tully-Fisher relation for rotating disk galaxies; they occupy a similar region 
in the Tully-Fisher plane as the HVCs and dwarfs, but cluster more tightly.  

\begin{figure}
\epsscale{1.0}
\plotone{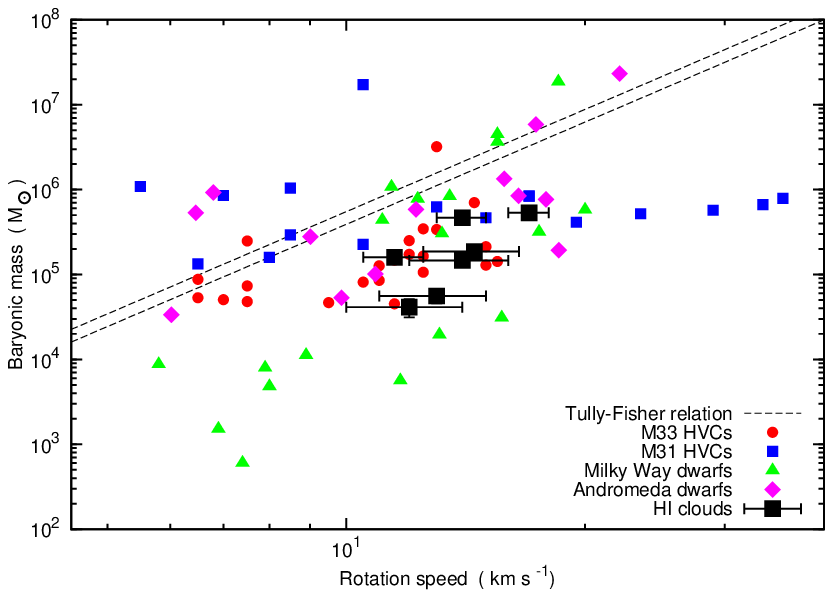}
\caption{{\bf A plot of baryonic mass versus equivalent circular rotation speed}. This figure compares the newly detected clouds (black filled squares with 1$\sigma$ errorbars), the HVCs of M31 and M33 (squares, circles), Milky Way dwarf satellite galaxies \citep{2010ApJ...722..248M} (triangles) and the Andromeda dwarfs \citep{2012ApJ...752...45T} (diamonds) to the baryonic Tully-Fisher relation \citep{2012AJ....143...40M} with $\pm 1 \sigma$ uncertainty (dashed lines). We assume that all of the mass is in the neutral atomic gas and that $\rm{M_{baryonic}=1.33 \times M_{HI}}$ to account for helium. The masses are plotted in units of solar masses. The presence of an ionized component will increase the baryonic mass and move the clouds closer to the relation, but we have no way at present to estimate its magnitude. The effective circular rotation speed is $\rm{v_{rot}=FWHM/2}$ where FWHM is the full width at half-maximum of the HI line. This assumption holds in the case where the gas may be in rotation or is a turbulently supported, spherical system with an assumed harmonic radius of $\rm{R/2}$, where R is the cloud radius, measured to same column density contour as the diameter in Table 1. The scatter in mass-rotation velocity is 0.5 dex for the clouds, 0.8 dex for the Andromeda dwarfs and M33's HVCs, 1.2 dex for M31's HVCs, and 1.8 dex for the Milky Way dwarfs. The average errors (in dex) for the rotation velocities $<\sigma_{v}> $ and the masses $<\sigma_{M}>$ are respectively 0.05 dex and 0.1 for M33's HVCs, 0.05 and 0.06 for M31's HVCs, 0.1 and 0.2 for the Milky Way dwarfs, 0.1 and 0.7 for the Andromeda dwarfs and 0.06 and 0.08 for the HI clouds.}
\end{figure}

If the linewidths of  the clouds are interpreted as an equilibrium velocity representative of  the gravitational potential, then the clouds would be highly dark matter dominated. The virial mass of every cloud is two to three orders of magnitude larger than its HI mass. The presence of an ionized component, however, would decrease the amount of dark matter necessary to match the virial mass. The free-fall time for these clouds, or time it would take for these clouds to freely collapse under their own gravity, is $\sim$ 400 Myr and the crossing time, the time it would take for a parcel of gas to cross the length of the cloud, is $\sim$ 100 Myr. These times are quite short to have resulted from the interaction between M31 and M33 proposed to have occurred a few billion years ago \citep{2008MNRAS.390L..24B, 2009Natur.461...66M} unless confined by an external pressure. The warm-hot intergalactic medium (WHIM) \citep{2001ApJ...552..473D} at a temperature $\sim 10^{6}$ K and density $\rm{\gtrsim 10^{-7}~cm^{-3}}$ would supply the pressure if the clouds have a temperature of $\sim 10^{4}$ K, consistent with the values inferred from the cloud linewidths. The WHIM densities needed are comparable to those predicted \citep{2004ApJ...616..643F} beyond the virialized region of galaxies, therefore these clouds could be in complete pressure equilibrium with the WHIM.

These clouds have several possible origins. First, they may be primordial, gas-rich objects similar to the dwarf spheroidal or dwarf irregular galaxies in the Local Group, but this does not explain why the gas seems to lie along a connecting structure, even when resolved into clouds. Second, the clouds may be gas that is accreting onto sub-halos. This might occur naturally if the gas is part of a cosmic filament. If instead the gas is tidal in origin, it is likely to be moving at relative speeds that exceed the escape velocities of sub-halos, making accretion unlikely. Third, the clouds may be analogous to tidal dwarf galaxies \citep{2007A&A...472L..25G}. This would account for their spatial and kinematical distribution with respect to M31 and M33. If so, the clouds would need to be in pressure equilibrium with the WHIM, so that they could persist for a few billion years following the interaction between M31 and M33. It's also possible that there may have been a more recent interaction that produced these clouds, such as M31 with IC10 or some other object. These clouds contain no evidence of stars, however, and do not fall on the Tully-Fisher relation, as one system of tidal dwarfs is known to do \citep{2007A&A...472L..25G}. Last, the clouds could be transient objects condensing from an intergalactic filament, as originally proposed \citep{2004A&A...417..421B, 2004MNRAS.355..694M}. This could explain their spatial distribution, as well as the lack of stellar overdensities. A recent simulation \citep{2012ApJ...749..181F} of a Milky Way sized galaxy has produced analogous material out to $\sim$ 100 kpc with HI masses $\sim 10^{5} ~\rm{M_{\odot}}$, suggesting that the clouds are most probably condensations from an intergalactic filament.

We conclude that the HI clouds between M31 and M33 are a unique population, lacking a clear analog in other members of the Local Group, such as the dwarf galaxies or HVCs. Although our current results suggest that the clouds may arise in an intergalactic filament, and thus are potential fuel to feed star formation in M31 and/or M33, the question of their origin needs further consideration. The new clouds we report here were discovered in only about half of our survey time and thus are not at the full sensitivity that we intend to achieve. The region we have studied is only a fraction of the area around M31 reported \citep{2004A&A...417..421B} to have diffuse HI, so the clouds measured here may be the first representatives of a much larger population.


 S.A.W. acknowledges the student observing support (GSSP11-012) provided by the NRAO for this project. The NRAO is a facility of the National Science Foundation operated under cooperative agreement by Associated Universities, Inc. D.J.P. acknowledges support from NSF CAREER grant AST-114949. S.S.M. acknowledges support in part by NSF grant AST-0908370.
All authors assisted in the development and writing of this work. S.A.W is the Principle Investigator, and was responsible for the data collection, processing and analysis. D.J.P. and F.J.L. assisted in the experimental design, data collection and analysis. S.S.M. provided the techniques of estimating the baryonic masses. E.J.S. provided necessary insights on potential interactions with M31 and ways of estimating circular rotation speeds. All authors aided in the interpretation of the results and provided comments for revisions to this work.
The authors declare no competing financial interests. Correspondence and requests for materials should be addressed to S.A.W. \ (email: swolfe4@mix.wvu.edu).


\bibliography{ArXBIB}

\newpage

\section*{Supplementary Figure}
Here we present channel maps of our GBT data (Supplementary Figure 1). The $9'$ spectral line data were gridded into a data-cube with the two spatial axes being right ascension and declination, while the third axis is the Local Standard of Rest (LSR) velocity in kilometers-per-second (seen in the upper left of each panel). The rms noise in the map ($1\sigma$) is $\sim 4$ mK. The dashed contour and first solid contour are for $\pm 2 \sigma$ respectively. The other solid contours are for $4\sigma$, $6\sigma$, $8\sigma$ and $10\sigma$. There is some low level striping evident in the cube. This is primarily due to certain areas of the map having a lower number of scans than others. This effect is only significant at the $2\sigma$ level. The maps from Figure 1 of the main text were generated by clipping the noise to emphasize emission above the $3\sigma$ level. Cloud 5 is visible from -197 to -223\kms. Clouds 2 and 3 become visible from -212 to -254\kms. Striping in the noise at -207\kms makes it difficult to determine if there is emission there as well. Cloud 2 is obviously extended in the northeast-southwest direction. Cloud 4 is visible from -254 to -310\kms and shows a possible extension to the southeast. Cloud 6 is seen from -269 to -295\kms. Cloud 1 is visible from -285\kms to -310\kms. Cloud 7 can be seen most clearly at -310 and -316\kms, but only slightly above $2\sigma$. This is why it is not visible in the $9'$ map, but is seen in the $15'$ map. This feature is unlikely due to striping since the emission persists over 5 channels, but the striping features only persist (locally) over one or two channels. 

\begin{figure*}
\epsscale{1.15}
\plotone{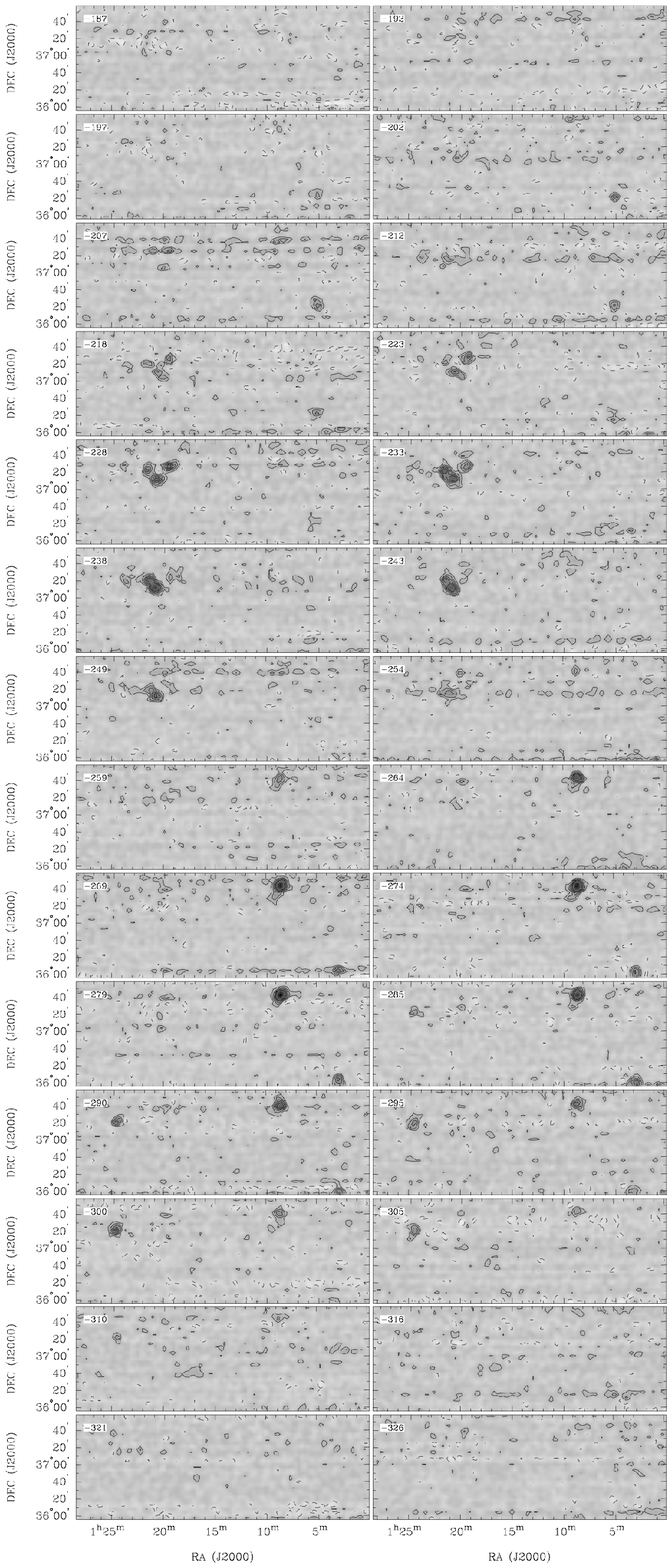}
\caption*{{\bf Supplementary Figure 1: Channel maps of the GBT data.}}
\end{figure*}

\end{document}